# ACTIVE VIBRATION SUPPRESSION R&D FOR THE NLC

J. Frisch, L. Hendrickson, T. Himel, A. Seryi, SLAC, Stanford, CA 94025, USA

*Abstract*

The nanometer scale beam sizes at the interaction point in linear colliders limit the allowable motion of the final focus magnets. We have constructed a prototype system to investigate the use of active vibration damping to control magnet motion. Inertial sensors are used to measure the position of a test mass, and a DSP based system provides feedback using electrostatic pushers. Simulation and experimental results for the control of a mechanically simple system are presented.

## 1 STABILIZATION REQUIREMENTS

The beam size at the interaction point of the NLC is approximately 2 X 200 nanometers (for 1 TeV CM system) [1,2]. In order to maintain luminosity, the relative positions of the beams at the IP must be stabilized to approximately 1 nanometer.

Vibration of the final focus quadrupoles is a potentially serious source of beam motion at the IP. Ground motion is one of the causes for this vibration. While ground motion in a quiet site could be sufficiently small, the "cultural" noise from man-made sources may substantially change the ground motion spectrum. Moreover, noises of the accelerator complex itself will contribute to this spectrum. Measurements at existing labs, showing 1.5nm of RMS motion above 1Hz at LEP [3] and 70nm at HERA [4], illustrate the magnitude of uncertainty of the initial conditions.

The NLC is being designed with careful engineering consideration of vibration issues. However, stabilization requirements for the final quadrupoles are hard to predict until the latest stage of design. Therefore, rather than to build a stabilization system to meet a specific requirement, work is underway to develop general stabilization technologies to be implemented at the final stage of NLC design.

## 2 STABILIZATION TECHNOLOGIES

*Beam Based Systems.* The interaction of the electron and positron beams at the IP causes a beam deflection that is related to the beam offset. This allows the offset to be measured to a fraction of the spot size at the beam rate of 120Hz. This beam deflection provides the only long-term measure of the relative positions of the beams.

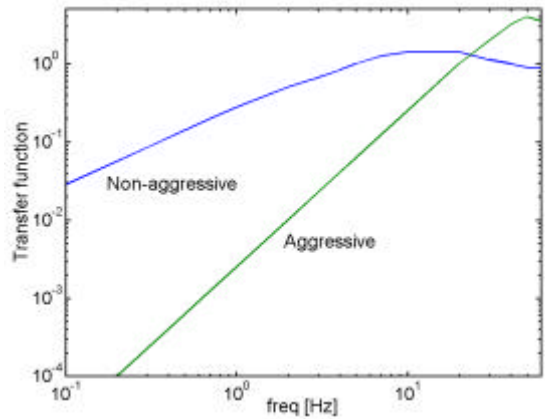

Figure 1: Simulated effect of beam-beam feedbacks.

A variety of feedback algorithms can be used with the beam – beam deflection data, with the selection based on the trade off between low frequency attenuation, and high frequency amplification of noise. Figure 1 shows the simulated effects of "aggressive" and "non-aggressive" beam feedbacks.

For the non-aggressive feedback the noise attenuation at 1Hz is 4 times, while the maximum amplification is 1.3 at 10-20Hz. For the aggressive feedback, the attenuation at 1Hz is 400 and the amplification at 50-60Hz is 4 times. Feedback algorithms with intermediate performance are also possible. One should note that the beam-beam feedback performance is ideal, i.e. it does not account for BPM resolution or jitter of beam size.

Very fast intra-train beam – beam feedback is also under consideration. This system relies on sufficient bandwidth and low latency to allow operation within the 400 nanoseconds NLC bunch train [5].

*Interferometer Based Systems.* Optical interferometers can be used to measure the distance between the final focus magnets and a fixed point or points on the ground. These "optical anchor" systems were proposed in the NLC ZDR [2] and are being developed at the University of British Columbia [6].

*Inertial Based Systems.* Inertial sensors can be used to measure the motion of the final quadrupoles relative to the "fixed stars". At low frequencies the position noise of an inertial sensor must rise sufficiently slow and a transition to a beam based system must be made.

A simple inertial vibration stabilization system has been constructed and is being used to test feedback hardware and algorithms.

# 3 VIBRATION FEEDBACK TEST SYSTEM DESIGN

*Mechanical Design.* An approximately 30kG aluminum block is suspended on 6 springs which define the 6 solid body degrees of freedom (see Figure 2). Six inertial sensors are used to provide feedback signals; a seventh measuring vertical motion is used to evaluate the feedback performance. Seven actuators, with two operated in parallel, are used to control the motions of the block.

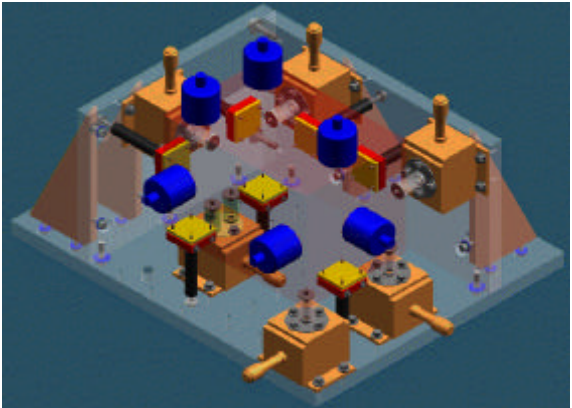

Figure 2: Drawing of the test system.

There is a trade-off on the support spring stiffness. High stiffness springs, which result in high normal mode frequencies, provide relatively low amplitude motion and good stability in the absence of feedback. They have the disadvantage of coupling high frequencies to the supported object, which are difficult to control with feedback.

Low stiffness springs allow large amplitude low frequency motions, but attenuate high frequencies. This approach was chosen for the test system, with normal mode frequencies in the range of 3-10Hz.

*Actuator Design.* For a system with a low stiffness support, the actuators must also have low stiffness. In order to close the feedback loop with high gain, the time delay of the actuators must also be short compared with other time constants in the system. While piezo-electric actuators are commonly used for this type of application, it is difficult for them to meet the above requirements.

We chose pure electrostatic actuators. A pair of electrodes, each 5x5cm, separated by approximately 1 mm, is used. With a maximum voltage of 1000V, the force produced is approximately 0.01 Newtons – sufficient to stabilize the mass motion. The response time is limited by the high voltage amplifiers (Trek 601C) which have a large signal bandwidth > 8 KHz.

*Sensors.* The test system uses commercial compact geophones (Geospace HS-1) with a 4.5Hz resonance and low noise (Analog Devices AD624 based) amplifiers. The resulting system has a (calculated) noise level of ~2nm/Hz$^{1/2}$ at 1Hz. The geophones are sensitive to the velocity of a test magnet inside of a pick-up coil, so their sensitivity decreases as $1/f^3$ below resonance.

In the NLC the final quadrupoles will be mounted inside the multi-Tesla detector solenoid, and will require non-magnetic inertial sensors. In addition, the noise level of the commercial geophones is currently limiting the performance of the test system.

New capacitive, non-magnetic accelerometers are being developed for this project. Electronics has been demonstrated with a sensitivity of 0.01nm/Hz$^{1/2}$ [7]. Mechanical design of a sensor is underway. This system is expected to be a factor of 50 lower noise at 1Hz, and 500 lower noise at 0.1Hz than the compact geophones currently in use.

*Data Acquisition System.* The data acquisition system is constructed from VME format hardware. A 16 bit 250KHz, 8 channel A-D / D–A card (Pentek 6102) is used to interface with the sensors and actuators. Closed loop feedback is performed with a TMS320C40 50MHz DSP (Pentek 4284) which interfaces to the A-D / D-A using a C40 port. This interface removes the real time communication from the VME backplane. A 68040-based crate controller (Motorola MVME167) running VxWorks is used for program loading and non-realtime data communications. The system is interfaced to Matlab for algorithm development.

# 4 TEST SYSTEM ALGORITHM

*Calibration and Orthogonalization.* Each of the 6 actuators is driven at each of 110 frequencies, while each of the sensor outputs is measured. The measurement frequencies include 50 logarithmically spaced from 1 - 30Hz, and 10 clustered near each of the resonance frequencies.

A 96 parameter fit is then done to the normal mode frequencies and Qs, sensor frequencies and Qs, actuator to mode couplings, and sensor to mode couplings. Figure 3 shows the quality of the fit. The resulting fit is tested by driving the combination of actuators corresponding to a single mode, and measuring the corresponding set of sensors. The lack of resonant peaks for the other modes indicates that the orthogonalization is good (Figure 4).

*Feedback algorithm.* The sensor measurements are converted to orthogonal modes. Then six independent feedback loops control vibration in each mode.

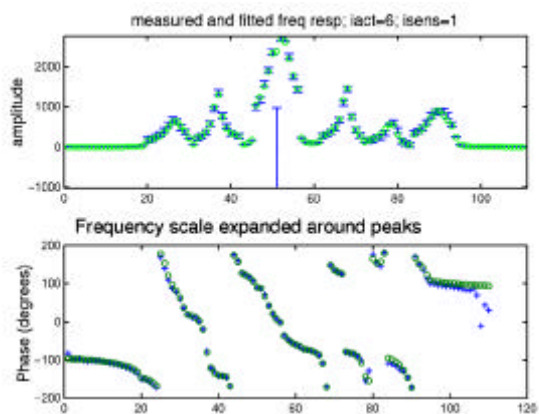

Figure 3: One of 36 measured calibration curves. The fit is shown by green circles.

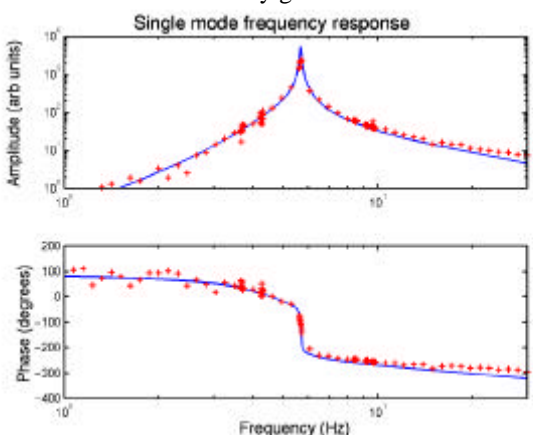

Figure 4: Frequency response of single mode. Measurements (symbols) and model.

The measured frequency response and noise spectral densities are used as inputs to generate an optimal feedback. Note that in the present system expected rather than measured noises are used, and identical mode frequency responses are assumed in order to simplify testing.

The feedback can be implemented as either state space, or transfer function (ZPK). State space is more numerically stable, but less computationally efficient. Results described are for state space running at a loop speed of 500Hz (not believed to be a limit on performance) with the gain set to 0.4 times optimal. The reduced gain is necessary to avoid instabilities. These instabilities are under study, but may be due to actuator saturation during system start-up.

*System Performance.* The test system performance is measured using the test sensor that measures vertical motion of the block. One should note that the test system is located in a fairly noisy lab environment – comparable to the "HERA" site, and that the noise of this test sensor becomes significant below ~1Hz.

The suspension of the block attenuates high frequencies, but introduces a low frequency resonance. The inertial feedback largely eliminates this resonance (see Figure 5).

In order to evaluate the performance of the system we can simulate the effect of beam-beam feedback on the measured output of the test sensor (lines with symbols in Figure 5). One can see that in a somewhat quieter place the aggressive beam feedback would already satisfy our requirements. The applicability of the aggressive feedback may however be limited by presence of uncorrelated jitter in the incoming beam.

Further improvements of the system performance are expected from optimization of the algorithm and in particular from new less noisy inertial sensors.

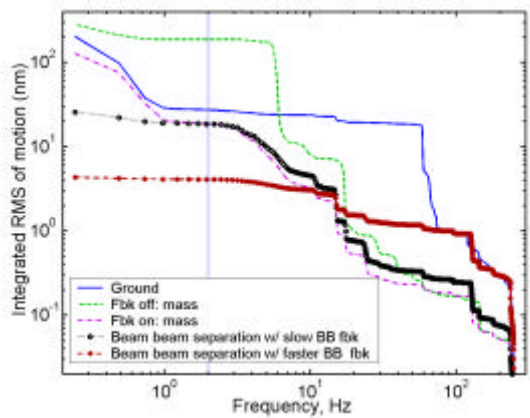

Figure 5: Integrated power spectrum showing performance of the test system.

## 5 FUTURE PLANS

The present system is mechanically simple, with the solid body normal mode frequencies of the block much lower than any internal resonant frequencies. When the operation of this system is well understood, an elongated structure with internal mode frequencies comparable to those of a real final focus magnet will be stabilized.